\documentclass{kluwer}
\usepackage[dvips]{graphicx}
\begin{document}
\begin{article}
\begin{opening}
\title{Gravitational Lensing}
\subtitle{A Universal Astrophysical Tool}

\author{Joachim \surname{Wambsganss}\email{jkw@astro.physik.uni-potsdam.de}}
\institute{Universit\"at Potsdam, 
		Institut f\"ur Physik, 
		Am Neuen Palais 10, 
		14469 Potsdam\\
and \\
		Max-Planck-Institut f\"ur Gravitationsphysik, 
		Am M\"uhlenberg 1,
		14476 Golm,
		Germany}




\runningtitle{Gravitational Lensing}
\runningauthor{Joachim Wambsganss}



\begin{abstract} 
In the roughly 20 years of its existence as an
observational science,
gravitational lensing has established itself as a valuable
tool in many astrophysical fields. 
In the introduction of this review we briefly present
the basics of lensing. 
Then it is shown that the two propagation effects,
lensing and scintillation, have a number of properties in common.
In the main part various lensing phenomena are discussed 
with emphasis on recent observations.

\end{abstract}

\keywords{astrophysics - cosmology - gravitational lensing - quasars}

%
%
\end{opening}
\section{The Basics of Lensing and Microlensing}

The general geometry of a gravitational lensing situation is shown in
Figure 1: the light path between observer O and source S
is affected by a mass L  between them.
In a ``strong" lensing situation -- 
lens and source are well aligned -- two (or more) images 
S$_1$ and S$_2$ of the background source can be produced. 
They are separated by an angle which is 
proportional to the square root of the lens mass (see below).
For typical galaxy masses this angle is of order arcseconds.

If the lens is a galaxy which consists 
partly of stars (and other compact objects), 
due to the graininess of the main lens, 
each of these (macro-)images consists of many
micro-images, which are separated by angles of order microarcseconds, 
and hence unresolvable. 
However, due to the relative motion between 
source, lens and observer, the micro-image configuration changes 
with time, and so does the observable total magnification.

From Figure 1, one can easily  derive the lens equation
(all angles involved are small: $\alpha, \tilde\alpha, \beta, \theta \ll 1$):
$$\vec{\beta} \times D_S = 
	\vec{\theta} \times D_S - \vec{\tilde{\alpha}} \times D_{LS}.$$
This relation reduces to $\vec{\beta} = \vec{\theta} - \vec{\alpha}$, 
where 
$$
	\vec{\alpha} = (D_{LS}/D_S) \times \vec{\tilde\alpha} = 
	(4 G M / c^2) (\vec\xi /      \xi^2) 
	\propto (\vec\theta/\theta^2)
$$
is the deflection angle for a point lens ($G$ - gravitational constant;
$c$ - speed of light; $\vec\xi$ - impact parameter of light ray;
$M$ -- mass of the lens; $D_L$, $D_S$, $D_{LS}$  --
angular diameter distances between observer-lens, observer-source, 
lens-source).
The positions of the images are the solutions of 
the quadratic equation for $\theta$.
For perfect alignment, i.e.  $\vec\beta = 0$,  the circular symmetric
image has a radius of 
$$ 
	\theta_E = 
	\sqrt{ { {4 G M } \over {c^2} } { {D_{LS} \over D_L D_S}  } }
		\approx    2 \sqrt {M / 10^{12} M_\odot} \ \ \rm arcsec
		\approx  2 \sqrt {M /M_\odot} \ \ \rm microarcsec. 
$$
This defines the relevant angular scale for gravitational lensing,
the angular Einstein radius $\theta_E$ of the lens
(``typical" lens and source redshifts of $z_L$$\approx$ 0.5 and
$z_S$$\approx$ 2.0  were assumed for the numerical values
on the right hand side).
%
%
%
%
%
%
%
\begin{figure}[htbp]
\centerline{\includegraphics[width=100mm]{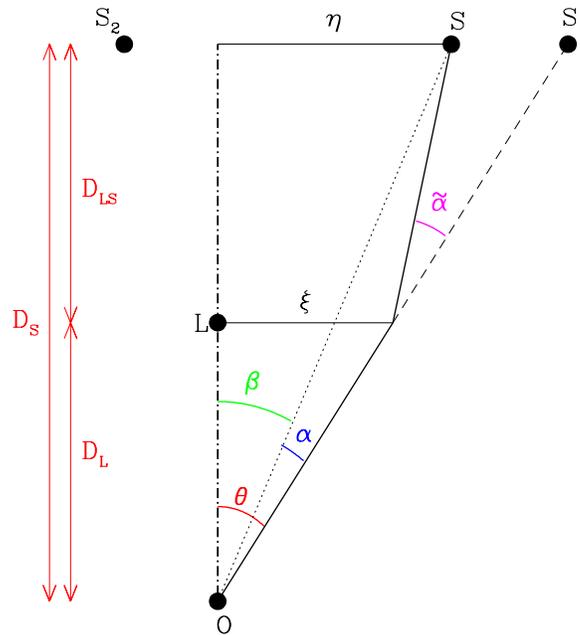}}
\caption{Basic setup of a gravitational lens situation. 
The lens  L 
in the lens plane 
produces two images S$_1$ and S$_2$ 
of a background source S.
If the lens has substructure (e.g., stars in a galaxy)
each macro-image 
is split into many micro-images. Only the
total magnification of all the microimages is observable. (The
symbols are explained in the text.) }
\label{fig1}
\end{figure}

\noindent
This angular scale translates into a length scale  in the source plane:
	$$ r_E = D_S \times \theta_E 
	\approx 4 \times 10^{16} \sqrt{M / M_\odot} \rm \,  cm 
	\approx        10  \sqrt{M / 10^{12} M_\odot} \rm \,  kpc.$$
The value on the right hand side
is typical for galaxy masses. This type of ``macro"-lensing
can be directly observed in form of multiply imaged quasars, 
with images separated by angles of order a few arcseconds.

Image splittings of microarcseconds corresponding to stellar mass lenses
can obviously not be detected directly. 
What makes ``microlensing" observable
anyway is the fact that observer, lens(es) and source move relative to each
other. Due to this relative motion, the micro-image configuration 
changes with time, and so does the total magnification, i.e. the
sum of the magnifications of all the micro-images. And this change
in magnification over time can be measured:
microlensing is a ``dynamical" phenomenon. 
There are two time scales involved: the standard lensing
time scale $t_E$ is the time it takes the source to cross the
Einstein radius of the lens, i.e. 

	$$ t_E = r_E/v_\perp   \approx  15 \sqrt {M / M_\odot}  v_{600}^{-1} \ \ \rm  years, $$
where the same typical assumptions are made as above, 
and the effective 
transverse velocity $v_{600}$ is parametrized in 
units of 600 km/sec. 
This time scale results in discouragingly large values. 
However, we can expect fluctations on much shorter time intervals, due
to the fact that the magnification distribution is highly non-linear, the
sharp caustic lines separate regions of low and high magnification. So
if a source crosses such a caustic line, we will observe a large
change in magnification during the time it takes the source to
cross its own diameter: 

	$$ t_{cross} 
= R_{source}/v_\perp   \approx   4 R_{15}  v_{600}^{-1} \ \ \rm  months.$$
Here the quasar size $R_{15}$ is parametrized in units of
$10^{15}$cm.
For a more detailled presentation of lensing,
see Schneider et al. (1992).

\section{Similarities between Lensing and Scintillation}

\begin{figure}[htbp]
\centerline{\includegraphics[width=120mm]{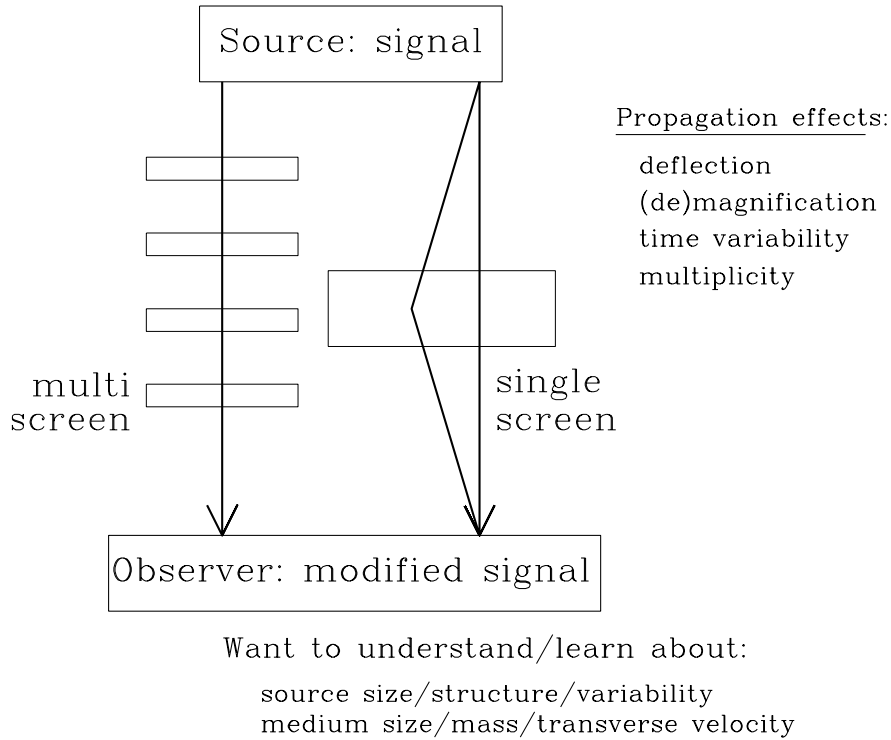}}
\caption{Similarities between gravitational lensing and
scintillation as propagation mechanisms}
\label{fig2}
\end{figure}

Gravitational lensing and interstellar scintillation are 
both propagation phenomena, and in fact these two physical 
effects have
quite a number of aspects in common. This is graphically
demonstrated in Figure \ref{fig2}.
In both cases, a signal emitted by a distant source is affected
by some intermediate medium, so that the observer receives
a modified signal.
The physical effect of the medium in both cases
is the {\it deflection of electromagnetic
		waves} (restricted wavelength range in the case 
		of scintillation), which causes 
\renewcommand{\labelenumi}{\alph{enumi})}
\begin{enumerate}
	\item shift of image position,
	\item apparent magnification or demagnification  of a source
		(relative to the unaffected situation),
	\item distorted image shape,
	\item time variability due to relative motion of source, 
		medium and observer, and
	\item occasionally multiplicity of images. 
\end{enumerate}
Both lensing and scintillation can be used  to
better understand intrinsic properties of the source:
size, structure, variability, as well as properties of the ``medium":
size, mass, transverse velocity.
The effect of the  medium  can be dominated by one relatively
strong interaction (single screen), or by a 
sum of many uncorrelated weak effects (multi screen).

\begin{table}[htbp]
\begin{tabular}{ l c }
\hline 
\ \ \ \ \ \ \ \ Phenomenon \ \ \ \ \ \ \ \ & \ \ \ \ \ \ \ \ Object of interest\ \ \ \ \ \ \ \  \\ 
\hline 
search for dark matter & medium \\
cosmology: $H_0$, $\Omega_{\sf matter}$, $\Omega_{\Lambda}$  & medium\\
size/structure/physics of quasars & source\\
galactic structure & medium \\
evolution of galaxies & source/medium \\
search for extra-solar planets & medium \\
\hline 
\end{tabular}
\caption{Goals of lensing and their relation to the propagation aspect
}
\label{tab1}
\end{table}

In Table \ref{tab1} a selection of  astrophysical fields 
in which gravitational lensing is important
is listed.
In order to stress the parallels to scintillation,
it is also indicated whether in this  particular field the
``object of desire" is the source or the medium.

\section{Lensing Phenomena}
In Table \ref{tab2} the most prominent gravitational lensing
phenomena are given with the approximate
number of detected cases and a recent
reference or review. 
A more detailed discussion of the various phenomena
with many examples can be found in 
Wambsganss (1998). An up-to-date 
version of the observational situation concerning
lensed quasars is available at the CASTLES
web page (Falco et al. 1999).

\begin{table}[htbp]
\begin{tabular}{ l c l }
\hline 
Phenomenon & \#of cases& Recent reference/review \\
\hline 
Multiple Quasars & $>$60   & CASTLES,  Falco et al. 1999 \\
Time Delays, $H_0$  & $\approx$ 6   & Myers     1999 \\
Giant Luminous Arcs & $\approx$100 & Mellier 1998 \\
Einstein rings   &  $\approx 10$     & CASTLES, Kochanek et al. 1999 \\
Quasar microlensing & $> 5$   & Wambsganss 2000 \\
``Arclets", weak lensing &  $\sim$ 200 & Mellier 1998 \\
Galactic microlensing & $ >  500$ & Alcock et al.  2000 \\
Lensing by large scale   & a few  & Bacon et al., Kaiser et al., van \\
structure/``cosmic shear" & & Waerbecke et al., Wittman et al. 2000\\
\hline 
\end{tabular}
\caption{Gravitational lens phenomena:  number of cases,
recent reference/review }
\label{tab2}
\end{table}

About 10\% of the roughly 60 known multiply imaged quasars have
a measured time delay.  Together with a model of the lensing galaxy,
this provides a determination of the Hubble constant. 
Myers (1999) has summarized the latest situation: 
values of the Hubble constant determined from gravitational
lensing tend to be lowish (around 65 km/s/Mpc).
Uncorrelated fluctuations in multiply imaged quasars are interpreted
as microlensing by stellar mass objects. It can be used to determine
the structure of quasars and the mass of the lensing objects.
(Wambsganss 2000).
 
Giant luminous arcs found in rich galaxy clusters
provide strong support of large masses for these clusters, which means
that they are dominated by dark matter. 
Even clusters at redshifts close to one still show  (weak) lensing
effects on background galaxies (Mellier 1999). 
 
Einstein rings -- annular images due to perfect alignment between 
lens and source -- are mostly known in the radio  regime.
Since they are ``extended", they provide many more constraints 
on the deflecting potential than a few point-images of quasars.
Hence the lensing galaxies of Einstein rings are probably those
with the most accurately determined  masses and  mass 
distributions (Kochanek, Keeton \& Mcleod 2000).
 
The most recent development of gravitational
lensing is the discovery of ``cosmic shear",
the very weak lensing effect of large scale structure on background
souces.  In a kind of stimulated emission, four groups presented
their uncorrelated detections within a few days this spring:
Bacon, Refregier \& Ellis (2000),
Kaiser, Wilson \& Luppino (2000),
van Waerbecke et al. (2000), and 
Wittman et al. (2000).

{}

\end{article}
\end{document}